\documentclass[aps, prd, nofootinbib, preprintnumbers,twocolumn, showpacs]{revtex4}
\usepackage{graphicx,color}
\usepackage{amsmath,amssymb}
\newcommand{\VEV}[1]{\langle #1 \rangle}

\newcommand{\diag}{\mathrm{diag}}

\def\fsl#1{\setbox0=\hbox{$#1$}                 % set a box for #1
   \dimen0=\wd0                                 % and get its size
   \setbox1=\hbox{/} \dimen1=\wd1               % get size of /
   \ifdim\dimen0>\dimen1                        % #1 is bigger
      \rlap{\hbox to \dimen0{\hfil/\hfil}}      % so center / in box
      #1                                        % and print #1
   \else                                        % / is bigger
      \rlap{\hbox to \dimen1{\hfil$#1$\hfil}}   % so center #1
      /                                         % and print /
   \fi}                                         %
\begin{document}
\title{Composite $Z'$}
\author{Michio Hashimoto}
\email{michioh@isc.chubu.ac.jp}
\affiliation{
   Chubu University, \\
   1200 Matsumoto-cho, 
   Kasugai-shi,  Aichi, 487-8501, JAPAN}
\date{\today}

\begin{abstract}
We investigate a possibility of a composite $Z'$ vector boson.
For the compositeness, the required gauge coupling $g$ in low energy
is not so big, $g^2/(4\pi) \gtrsim 0.015$ in the case of
the $U(1)_{B-L}$ model. 
We show that the St\"{u}ckelberg model is effectively induced
in low energy via the fermion loop from the Nambu-Jona-Lasinio (NJL) model 
having the vectorial four-fermion interaction.
In terms of the renormalization group equations (RGE's), 
this situation is expressed by the compositeness conditions.
We find that the solutions of the RGE's with the compositeness conditions
are determined by the infrared fixed points.
As a result, the ratio of the masses of the extra electroweak singlet scalar
and the right-handed neutrino is fixed.
The mass of the composite $Z'$ boson contains 
the contribution $\Delta$ of the St\"{u}ckelberg mass term.
This nonzero $\Delta$ might be a remnant of a strongly interacting theory
in high energy.
\end{abstract}
\maketitle

\section{Introduction}

Recently, the Higgs boson has been discovered at the LHC~\cite{Higgs-discover}.
The Higgs mass is revealed around 125~GeV and the nature seems to be consistent
with the standard model (SM)~\cite{global-Higgs}.
This suggests that the magnitude of the Higgs quartic coupling $\lambda_H$ is 
$\lambda_H \sim 0.1$.
It then turns out that the perturbation works up to
very high energy scale, although $\lambda_H$ becomes negative 
at some scale~\cite{stability}.
Recall that the color and weak gauge interactions are asymptotic free
within the SM.
The SM top-Yukawa coupling $y_t$ is also decreasing function 
with respect to the energy scale $\mu$.
The hypercharge gauge coupling $g_Y$ 
is increasing, but it stays perturbative up to the Planck scale $\Lambda_{Pl}$.
Therefore it seems that there is no room for strongly interacting theories
in high energy. 

On the other hand, 
the gravity is apparently strong around the Planck scale.
If the complicated Yukawa structure of the SM is dynamically generated 
as in low energy QCD, we may expect some strong dynamics in a high energy
scale~\cite{Dimopoulos:1979es,Eichten:1979ah}.
Is there a remnant of such a strong interaction?

Nonzero neutrino mass is one of the evidences of physics beyond the SM.
Considering the quark--lepton 
correspondence~\cite{Katayama:1962mx,Maki:1962mu,Yasue:1978wg,Davidson:1978pm},
the existence of the right-handed neutrinos is highly possible.
The see-saw mechanism might help to generate tiny neutrino masses~\cite{seesaw}.
Since the Majorana mass term explicitly breaks the lepton number,
we may introduce an extra complex scalar field and a Majorana Yukawa 
coupling between the scalar and the right-handed neutrino.
A simple extension of the SM including the above is 
the $B-L$ model~\cite{Langacker:2008yv,Leike:1998wr}.
The right-handed neutrino is now {\it inevitably} required 
in order to cancel the anomaly of the $U(1)_{B-L}$ gauge current.

In this paper, we explore the possibility of a composite $Z'$.
Note that we do not need so strong $U(1)$ gauge coupling, 
if the strong coupling region is around the Planck or the GUT scale.

The Landau pole $\Lambda$ of the $U(1)$ gauge interaction is expressed as
$\Lambda = M \exp (8\pi^2/a g^2)$, where $g$ and $a$ denote
the $U(1)$ gauge coupling and the coefficient of the renormalization
group equation (RGE) for $g$, respectively.
The typical mass scale of the $U(1)$ gauge boson is $M$.
In the case of $U(1)_Y$, the Landau pole is well above the Planck scale,
$\Lambda \sim 10^{42}$ GeV by taking $M \sim 100$~GeV, 
$a=41/6$ and $g_Y^2/(4\pi) \simeq 0.01$.
This is the reason why the $U(1)_Y$ gauge interaction stays perturbative.
However, in order to obtain $\Lambda \lesssim \Lambda_{Pl}$,
we just need $g^2/(4\pi) \gtrsim 0.015$
for the $U(1)_{B-L}$ model, where $a=12$~\cite{Basso:2010jm}.
This lower bound is about half of the weak coupling square,
$g_2^2/(4\pi) \simeq 0.03$. 
Of course, bigger $U(1)_{B-L}$ gauge coupling implies 
an exponentially smaller scale of the Landau pole.
In any case, if $g^2/(4\pi) \gtrsim 0.015$ for the $U(1)_{B-L}$ model
is established in experiments, this suggests the existence 
of the strong interaction in high energy.
If so, we may expect that the $Z'$ boson associated with 
the $U(1)_{B-L}$ model is composite. 
(For the earlier approach, see, e.g., Ref.~\cite{Terazawa:1976xx}.)
%where the SM gauge bosons are composite.)
% (See also the subquark model where the SM gauge bosons and 
% the Higgs scalar as well as leptons and quarks are 
% all composite~\cite{Terazawa:1976xx,Terazawa:1979pj}.) 

First, we show that
the strong gauge coupling limit of the St\"{u}ckelberg 
model~\cite{Stueckelberg:1900zz,Ruegg:2003ps} 
corresponds to the Nambu-Jona-Lasinio (NJL) model with 
the vectorial four-fermion interaction. 
Next, we argue that the St\"{u}ckelberg model 
is effectively induced in low energy via the bubble diagram.
For the conceptual diagrams, see Figs.~\ref{prop-s} and \ref{prop-g}.
The compositeness at the scale $\Lambda$ can be described 
by the compositeness conditions,
\begin{equation}
  \frac{1}{g^2(\Lambda)} = \frac{1}{y^2(\Lambda)}=0, \quad
  \frac{\lambda(\Lambda)}{y^4(\Lambda)}=0,
\end{equation}
where $y$ and $\lambda$ are the Majorana Yukawa interaction
and the quartic coupling of the extra complex scalar field, respectively,
as in 
the Bardeen, Hill and Lindner (BHL) approach~\cite{Bardeen:1989ds}
of the top condensate model~\cite{Miransky:1988xi,Nambu,Hill:2002ap}. 
(See also the earlier attempt~\cite{Terazawa:1990mz}.)
The behavior of $y^2/g^2$ is controlled by the Pendleton--Ross type
infrared fixed point (PR-IRFP)~\cite{Pendleton:1980as}. 
Also, we find that $\lambda/g^2$ is determined by an infrared fixed point.
In our approach, the ratio of the masses of the extra scalar $\chi$ and 
the right-handed neutrino $\nu_R$ is fixed to $M_\chi/M_{\nu_R} \approx 1.2$.
In sharp contrast to the conventional $U(1)_{B-L}$ model, 
the $Z'$ mass has the contribution of the St\"{u}ckelberg mass term, so that 
$\Delta \equiv M_{Z'}^2/g^2 - 4v_\chi^2 > 0$, 
where $\VEV{\chi} = v_\chi/\sqrt{2}$.
This nonzero $\Delta$ might be the remnant of the strong dynamics 
in high energy.

Although we study only a fine-tuning scenario with a light composite $Z'$ 
in this paper, it might be more natural that the masses of $Z'$, $\chi$ and 
$\nu_R$ are around the compositeness scale $\Lambda$.
In the latter case, we will use the see-saw mechanism~\cite{seesaw}. 
We here mention that the Higgs potential can be stabilized owing to
the tree level threshold corrections for the Higgs quartic coupling,
which is generated by the $Z'$ loop effect~\cite{EliasMiro:2012ay}. 
We will investigate such a possibility elsewhere.

\section{St\"{u}ckelberg formalism and composite vector field}
\label{sec2}

\subsection{Strong coupling limit of the St\"{u}ckelberg model}
\label{sec2-1}

Let us first revisit the St\"{u}ckelberg formalism for 
the massive photon~\cite{Ruegg:2003ps,Langacker:2008yv}.
Introducing the St\"{u}ckelberg scalar field $B$,
the Lagrangian density of a massive vector field $A_\mu$ is 
\begin{equation}
{\cal L} = {\cal L}_\psi + {\cal L}_g + {\cal L}_{\rm gf},
\end{equation}
with 
\begin{equation}
 {\cal L}_\psi = \bar{\psi} i\fsl{\partial}\psi + g \bar{\psi} \fsl{A} \psi,
%  - M_\psi \bar{\psi} \psi, 
  \label{Lf}
\end{equation}
\begin{equation}
  {\cal L}_g = - \frac{1}{4} F_{\mu\nu} F^{\mu\nu}
  + \frac{1}{2} g^2 f^2 \left(A_\mu - \frac{1}{gf} \partial_\mu B\right)^2,
  \label{Lg}
\end{equation}
and
\begin{equation}
  {\cal L}_{\rm gf} = - \frac{1}{2} \left(\partial_\mu A^\mu + gf B\right)^2 , 
  \label{Lgf}
\end{equation}
where $\psi$ is a four-component fermion,
$F_{\mu\nu}$ denotes the field strength, 
\begin{equation}
  F_{\mu\nu} \equiv \partial_\mu A_\nu - \partial_\nu A_\mu ,
\end{equation}
and we represented the mass of the vector field $A_\mu$ by 
a gauge coupling constant $g$ times a parameter $f$ having mass-dimension one.
Although $\psi$ may have a Dirac mass, it is irrelevant in the following
discussions.
The parts of ${\cal L}_\psi$ and ${\cal L}_g$ are invariant under 
the (infinitesimal) local $U(1)$ transformation,
\begin{eqnarray}
  \delta A_\mu &=& \partial_\mu \omega (x), \quad
  \delta B = gf \omega(x),  \\
  \delta \psi &=& ig \omega \psi, \qquad
  \delta \bar{\psi} = -ig \omega \bar{\psi} \, .
\end{eqnarray}
Unlike the Proca field, this system is power-counting renormalizable.
We also note that we can rewrite the gauge fixing term 
by using the Becchi-Rouet-Stora-Tyutin (BRST) construction.\footnote{
The St\"{u}ckelberg scalar field $B$ has a positive metric,
in sharp contrast to the Nakanishi-Lautrup field,
which is introduced in a general BRST invariant construction of 
the gauge fixing term.
}

We may normalize $A_\mu$ and $B$ as $g A_\mu \to A_\mu$ and $gB \to B$, and
then we read the gauge sector,
\begin{equation}
  {\cal L}_g \to  - \frac{1}{4g^2} F_{\mu\nu} F^{\mu\nu}
  + \frac{1}{2} f^2 \left(A_\mu - \frac{1}{gf} \partial_\mu B\right)^2,
\end{equation}
also the gauge fixing term  becomes
\begin{equation}
  {\cal L}_{\rm gf} \to - \frac{1}{2g^2} (\partial_\mu A^\mu + g f B)^2 \,.
\end{equation}
Formally, taking the limit of $g \to \infty$, we obtain
\begin{equation}
  {\cal L} = \bar{\psi} i\fsl{\partial}\psi + \bar{\psi} \fsl{A} \psi
%  - M_\psi \bar{\psi} \psi 
  + \frac{1}{2} f^2 A_\mu^2  - \frac{1}{2} f^2 B^2,
\end{equation}
where we dropped the total derivative term.
The St\"{u}ckelberg fields $A_\mu$ and $B$ are now the auxiliary ones 
and thus, by performing the Gauss integral, we find
\begin{equation}
  {\cal L} \to \bar{\psi} i\fsl{\partial}\psi
  - G_V (\bar{\psi}\gamma^\mu \psi)^2 , %- M_\psi \bar{\psi} \psi ,
\end{equation}
with $G_V = 1/(2f^2)$ and $A^\mu \sim \bar{\psi} \gamma^\mu \psi$.
That is the NJL model with the vectorial four-fermion interaction 
is equivalent to the strong coupling limit of the St\"{u}ckelberg model.

\subsection{St\"{u}ckelberg model as a low energy effective theory 
and compositeness conditions}
\label{sec2-2}

\begin{figure}[t]
  \begin{center}
  \resizebox{0.35\textheight}{!}
            {\includegraphics{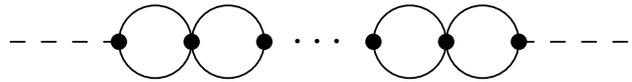}}
  \end{center}
  \caption{Conceptual diagram of the induced propagator for $\chi$
  by the scalar four-fermion interaction represented by the black dots.
  \label{prop-s}}
\end{figure}

\begin{figure}[t]
  \begin{center}
  \resizebox{0.35\textheight}{!}
            {\includegraphics{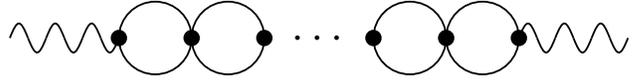}}
  \end{center}
  \caption{Conceptual diagram of the induced propagator for $A_\mu$
  by the vector four-fermion interaction represented by the black dots.
  \label{prop-g}}
\end{figure}

We now start from a model with a Majorana-type scalar four-fermion coupling
and a vector one:
\begin{equation}
  {\cal L} = \bar{\eta} i \fsl{\partial} \eta
  + G_S (\overline{\eta^c} \eta)(\overline{\eta} \eta^c)
  - G_V (\overline{\eta}\gamma^\mu \eta)^2,
  \label{NJL-Maj}
\end{equation}
where $\eta$ is a two-component fermion,
for example, a right-handed neutrino, and
$\eta^c$ is the charge conjugation.
This system classically has a global $U(1)$ symmetry,
\begin{equation}
  \eta \to e^{i\theta} \eta, \quad
  \overline{\eta} \to e^{-i\theta} \overline{\eta} \, .
\end{equation}
By introducing auxiliary fields,
$\phi \sim \overline{\eta} \eta^c$,
$\phi^\dagger \sim \overline{\eta^c} \eta$,
and $A_\mu \sim \overline{\eta}\gamma^\mu \eta$,
we can rewrite the model as follows:
\begin{eqnarray}
  {\cal L} &\to& \bar{\eta} i \fsl{\partial} \eta
  - \overline{\eta^c} \eta \phi 
  - \overline{\eta} \eta^c \phi^\dagger
  - M_{\phi,0}^2 \phi^\dagger \phi \\
&&
  + \overline{\eta}\gamma^\mu \eta A_\mu
  + \frac{1}{2} f_0^2 A_\mu^2 ,
\end{eqnarray}
with $M_{\phi,0}^2=1/G_S$ and $f_0^2=1/(2G_V)$.

In low energy,
the composite scalar and vector fields acquire the kinetic terms
via the bubble diagrams,\footnote{
Although we here introduced only a two-component fermion just for 
a pedagogical explanation,
in a realistic $Z'$ model discussed later,
there are many fermions: 
Three $\nu_R$'s contribute to both Figs.~\ref{prop-s} and \ref{prop-g}, 
and also the SM fermions contribute to Fig.~\ref{prop-g}, because they have
the $B-L$ charges. 
Thus the $1/N$-approximation is applicable.
}
similar to the top condensate model
\`{a} la BHL~\cite{Bardeen:1989ds}. 
See also Figs.~\ref{prop-s} and \ref{prop-g}.
The calculation for the vector-vector correlator is rather tricky, however.
As is well-known, a sharp cutoff regularization, which is usually used in 
the calculation of the two point function of the scalar sector,
is not appropriate.
Instead, we employ the proper time regularization.
The vacuum polarization diagram then yields
\begin{equation}
  \Pi^{\mu\nu}(p) = -\frac{1}{24\pi^2} (p^2 g^{\mu\nu} - p^\mu p^\nu)
  \log \Lambda^2,
\end{equation}
where we introduced the cutoff $1/\Lambda^2$ in the infrared part of 
the proper time integral.
Thus the induced effective theory in a low energy scale $\mu$ is
\begin{eqnarray}
 {\cal L}_{\rm eff} &=& \bar{\eta} i\fsl{D}\eta
%  + \bar{\eta} \gamma^\mu \eta A_\mu
  + Z_\phi |D_\mu \phi|^2 - M_\phi^2 \phi^\dagger \phi 
  - \lambda_\phi (\phi^\dagger \phi)^2 \nonumber \\
&&
  - \overline{\eta^c} \eta \phi 
  - \overline{\eta} \eta^c \phi^\dagger 
  - \frac{Z_A}{4} F_{\mu\nu} F^{\mu\nu}
  + \frac{1}{2} f^2 A_\mu^2,
\end{eqnarray}
where $D_\mu \eta = \partial_\mu \eta - i A_\mu \eta$, 
$D_\mu \phi = \partial_\mu \phi + 2i A_\mu \phi$,
and the scalar quartic coupling $\lambda_\phi$ is induced 
by the bubble diagram.
The wave function renormalization constants are
\begin{equation}
  Z_\phi = \frac{1}{16\pi^2} \log \Lambda^2/\mu^2, \quad
  Z_A = \frac{1}{24\pi^2} \log \Lambda^2/\mu^2 \, .
\end{equation}
Let us introduce 
\begin{equation}
  g \equiv \frac{1}{Z_A^{\frac{1}{2}}}, \quad 
  y \equiv \frac{1}{Z_\phi^{\frac{1}{2}}},
\end{equation}
and rescale $A_\mu$ and $\phi$ as $A_\mu \to g A_\mu$ and $\phi \to y\phi$,
respectively.
The effective theory thereby has the canonical kinetic terms, 
\begin{eqnarray}
 {\cal L}_{\rm eff} &=& \bar{\eta} i\fsl{D}\eta
  + |D_\mu \phi|^2 - \tilde{M}_\phi^2 \phi^\dagger \phi 
  - \tilde{\lambda}_\phi (\phi^\dagger \phi)^2 \nonumber \\
&& \hspace*{-5mm}
  - y \overline{\eta^c} \eta \phi 
  - y \overline{\eta} \eta^c \phi^\dagger 
  - \frac{1}{4} F_{\mu\nu} F^{\mu\nu}
  + \frac{1}{2} g^2 f^2 A_\mu^2 \, .
\end{eqnarray}
We further carry out field-dependent rotations for
the fermion and scalar variables,
\begin{equation}
  \varphi \equiv e^{i\frac{B(x)}{gf}} \eta, \quad
  \overline{\varphi} \equiv e^{-i\frac{B(x)}{gf}} \overline{\eta},
\end{equation}
and
\begin{equation}
  \chi \equiv e^{-2i\frac{B(x)}{gf}} \phi, \quad
  \chi^\dagger \equiv e^{2i\frac{B(x)}{gf}} \phi^\dagger, 
\end{equation}
also redefine the gauge field 
$\tilde{A}_\mu \equiv A_\mu + \frac{1}{gf} \partial_\mu B$.
Because we introduced a redundant field $B(x)$,
we should add a delta function $\delta (\xi_B -1)$ with
$\xi_B \equiv e^{i\frac{B(x)}{gf}}$ in the path integral.
Although the fermion path integral measure yields 
anomaly from these rotations,  
this is because we considered a simple model just
for a schematic explanation.
In the next section, we study an anomaly-free theory.
Any vectorial four-fermion couplings consistent with the symmetry
would be possible at the compositeness scale $\Lambda$, but, 
only the anomaly-free combinations among the composite gauge interactions
should be left in low energy.

In any case, the theory is then
\begin{eqnarray}
 {\cal L}_{\rm eff} &=&
    \bar{\varphi} (i\fsl{\partial} + g \fsl{\tilde{A}}) \varphi
  + |(\partial_\mu + 2i g \tilde{A}_\mu) \chi|^2
  - \tilde{M}_\chi^2 \chi^\dagger \chi \nonumber \\
&&
  - \lambda (\chi^\dagger \chi)^2 
  - y \overline{\varphi^c} \varphi \chi 
  - y \overline{\varphi} \varphi^c \chi^\dagger \nonumber \\
&&
  - \frac{1}{4} F_{\mu\nu} F^{\mu\nu}
  + \frac{1}{2} g^2 f^2
    \left(\tilde{A}_\mu - \frac{1}{gf} \partial_\mu B\right)^2 ,
    \label{Leff}
\end{eqnarray}
which is essentially equivalent to 
the St\"{u}ckelberg model with the complex scalar field.
The gauge fixing term comes from the delta function $\delta (\xi_B -1)$
added in the partition function.
In the NJL picture, we cannot avoid quadratically fine-tuning
in order to obtain wanted values of the mass terms for the composite 
scalar and vector fields in (\ref{Leff}) from (\ref{NJL-Maj}).

In summary, we introduced the redundant field $B(x)$,
which corresponds to the St\"{u}ckelberg scalar field,
and thereby the global $U(1)$ symmetry is upgraded to the local one,
\begin{eqnarray}
  \varphi &\to& e^{ig\omega(x)} \varphi, \quad
  \overline{\varphi} \to e^{-ig\omega(x)} \overline{\varphi} , \\
  \chi &\to& e^{-2ig\omega(x)} \chi, \quad
  \chi^\dagger \to e^{2ig\omega(x)} \chi^\dagger, \\
  \tilde{A}_\mu &\to& \tilde{A}_\mu + \partial_\mu \omega, \quad
  B \to B + gf \omega(x) \, .
\end{eqnarray}
In the hidden local symmetry approach for the chiral symmetry breaking,
the polar decomposition of the complex scalar field is 
used~\cite{Bando:1987br,Harada:2003jx}.
The hidden symmetry is essentially connected with the ambiguity
of the polar decomposition.
If we apply the same manner in the above model, 
the $U(1)$ symmetry is {\it explicitly} broken down
owing to the Majorana-type interaction.
In our approach, when the $\chi$ field develops a nonzero vacuum
expectation value (VEV) in some low energy scale,
the local $U(1)$ symmetry is {\it spontaneously} broken down.
In this case, the gauge field acquires the mass from both 
the nonzero VEV of $\chi$ and the St\"{u}ckelberg mass term built 
in the model from the very beginning.

Furthermore, it turns out that 
the St\"{u}ckelberg model as a low energy effective theory
corresponds to the composite model in a high energy scale $\Lambda$, 
when we impose the compositeness conditions,
\begin{equation}
  \frac{1}{g^2 (\Lambda)} = \frac{1}{y^2 (\Lambda)} = 0, \quad
  \frac{\lambda (\Lambda)}{y^4 (\Lambda)} = 0 \, .
\end{equation}

In the next section, we apply this formulation to a realistic $Z'$ model. 

\section{Composite $Z'$ model}

Let us study the $U(1)_{B-L}$ extension of the SM.
The Lagrangian density is 
\begin{equation}
  {\cal L} = {\cal L}_{\rm SM} + {\cal L}_{\nu} + {\cal L}_\chi
  + {\cal L}_{Z'} + {\cal L}_{\rm gf},
\end{equation}
where ${\cal L}_{\rm SM}$ represents the SM part, and
\begin{equation}
  {\cal L}_\nu = \sum_{f=1,2,3} \overline{\nu_R^f} i\fsl{D} \nu_R^f
\end{equation}
\begin{eqnarray}
  {\cal L}_\chi &=& |D_\mu \chi|^2 - M_\chi^2 \chi^\dagger \chi
  - \lambda_\chi (\chi^\dagger \chi)^2 - \lambda_{\chi H} |H|^2 |\chi|^2
  \nonumber \\
&&
  - Y_{jk} \overline{\nu_R^{j\,c}}\nu_R^k \chi
  - Y_{jk} \overline{\nu_R^j}\nu_R^{k\,c} \chi^\dagger, 
\end{eqnarray}
\begin{equation}
  {\cal L}_{Z'} = - \frac{1}{4} F_{\mu\nu} F^{\mu\nu}
  + \frac{1}{2} g^2 f^2 \left(A_\mu - \frac{1}{gf} \partial_\mu B\right)^2\,.
  \label{Zp}
\end{equation}
The field $H$ and ${\cal L}_{\rm gf}$ denote the SM Higgs doublet and
the gauge fixing term, respectively.
The $U(1)$ part of the covariant derivative is 
\begin{eqnarray}
  D_\mu &=& \partial_\mu 
  - i Q_Y (g_Y Y_\mu + \tilde{g} A_\mu)
  - i g Q_{B-L} A_{\mu}, 
\end{eqnarray}
where $Q_Y$ and $Q_{B-L}$ denote the hypercharge and the $B-L$ charge, 
respectively.
The gauge couplings of $U(1)_Y$ and $U(1)_{B-L}$ 
are $g_Y$ and $g$, respectively.
In general, the gauge mixing coupling $\tilde{g}$ appears.
It is natural to set $\tilde{g}(\Lambda) = 0$, because of 
no gauge kinetic mixing term at the compositeness scale $\Lambda$.
As for the scalar quartic mixing $\lambda_{\chi H}$, 
the operator $|H|^2 |\chi|^2$ has a higher dimension than six
at the compositeness scale $\Lambda$.
Thus we may safely neglect it, i.e., $\lambda_{\chi H}(\Lambda)=0$. 

Unlike the conventional $Z'$ model, 
the St\"{u}ckelberg mass term is incorporated in 
this composite model as in Eq.~(\ref{Zp}).
This 
is essential in our formalism of the composite vector field.

The full set of the RGE's for the $U(1)_{B-L}$ model is
shown in Refs.~\cite{Iso:2009ss,Basso:2010jm,Hashimoto:2014ela}.
Essentially, the RGE's for the gauge and Yukawa couplings are
\begin{eqnarray}
  \beta_g &\equiv& \mu \frac{\partial }{\partial \mu} g =
  \frac{a}{16 \pi^2}  g^3, \\
  \beta_y &\equiv& \mu \frac{\partial }{\partial \mu} y =
  \frac{y}{16 \pi^2} \bigg[\, b y^2 - c g^2\,\bigg], 
\end{eqnarray}
where we took $Y_{jk} = \diag (y,y,y)$.
The coefficients of the RGE's are $a=12$, $b=10$, and $c=6$.
We may take the number of the right-handed neutrinos having relevant 
Majorana yukawa couplings to $N_\nu$ in general, and 
then the coefficient $b$ is $b=4+2N_\nu$.
For details, see Refs.\cite{Hashimoto:2013hta,Hashimoto:2014ela}. 
We fix $N_\nu=3$ hereafter.
Imposing the compositeness conditions, $1/g^2(\Lambda)=1/y^2(\Lambda)=0$,
we analytically find the solutions,
\begin{equation}
  \frac{1}{g^2(\mu)} = \frac{a}{8\pi^2} \ln \frac{\Lambda}{\mu}, \quad
  \frac{1}{y^2(\mu)} = \frac{b}{a+c}\,\frac{1}{g^2 (\mu)} ,
  \label{PRFP}
\end{equation}
where $\Lambda$ is the compositeness scale.
This solution corresponds to 
the PR-IRFP~\cite{Pendleton:1980as}\footnote{
Strictly speaking, the asymptotic free theory, i.e., $a < 0$,
is used in Ref.~\cite{Pendleton:1980as}. 
Thus the situation $1/g^2(\Lambda) \to 0$ occurs in low energy
unlike in the asymptotic nonfree theory.
}.
Actually, we can easily rewrite the RGE's as follows:
 \begin{equation}
  (8\pi^2)\,\mu \frac{\partial }{\partial \mu} \left(\frac{y^2}{g^2}\right)
  = b \, g^2 \cdot \frac{y^2}{g^2}
    \left(\,\dfrac{~y^2~}{g^2} - \frac{a+c}{b}\,\right),
  \label{RGE-PR}
\end{equation}
and hence the solution (\ref{PRFP}) is the PR-IRFP.
Owing to this nature of the PR-IRFP, if we relax the compositeness conditions
to $1/g^2(\Lambda), 1/y^2(\Lambda) \ll 1$ (nonvanishing),
the RG flows are not changed so much. 
The RGE for $\lambda_\chi$ is a bit complicated:
\begin{eqnarray}
    \beta_{\lambda_\chi} &\equiv&
    \mu \frac{\partial }{\partial \mu} \lambda_\chi =
    \frac{1}{16 \pi^2} \bigg[\, 20 \lambda_\chi^2
    + \lambda_\chi (24 y^2 - 48 g^2) \nonumber \\
&& \qquad \qquad - 48 y^4 + 96 g^4\,\bigg], 
\end{eqnarray}
where we ignored the numerically irrelevant $\lambda_{\chi H}^2$ term.
Imposing the compositeness condition for $\lambda_\chi$,
$\lambda_\chi(\Lambda)/y^4(\Lambda) = 0$, and
positivity of $\lambda_\chi$ in any scale,
we obtain the analytical solution, 
\begin{equation}
  \lambda_\chi (\mu) = \frac{2}{25} \big(9+\sqrt{546}\big) \,g^2(\mu) \,.
  \label{sol-lam}
\end{equation}
This is a new infrared fixed point (IRFP) solution.
Substituting the solutions (\ref{PRFP}) for $g$ and $y$, 
we find
\begin{eqnarray}
  (16\pi^2) \mu \frac{\partial }{\partial \mu}
  \left(\frac{\lambda_\chi}{g^2}\right) 
 = 20 g^2 \left(\frac{\lambda_\chi}{g^2} - k_+ \right)
     \left(\frac{\lambda_\chi}{g^2} - k_- \right),
\end{eqnarray}
where $k_+ \equiv \frac{2}{25} \big(9+\sqrt{546}\big) \simeq 2.589$ and 
$k_- \equiv \frac{2}{25} \big(9-\sqrt{546}\big) \simeq -1.149$.
Thus the analytical solution (\ref{sol-lam}) corresponds to
the IRFP in fact.

\begin{figure}[t]
  \begin{center}
  \resizebox{0.35\textheight}{!}
            {\includegraphics{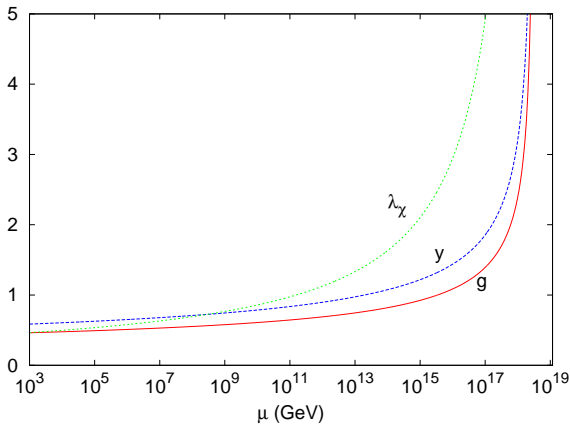}}
  \end{center}
  \caption{The RG flows of $g$, $y$ and $\lambda_\chi$.
  We took $\Lambda = 1/\sqrt{8\pi G} = 2.435 \times 10^{18}$ GeV, and
  imposed the compositeness conditions, $1/g^2(\Lambda)=1/y^2(\Lambda)=0$
  and $\lambda(\Lambda)/y^4(\Lambda)=0$.
  \label{flows}}
\end{figure}

We depict the RG flows in Fig.~\ref{flows}, by using the full set of
the RGE's. 
We took $m_t=173.5$~GeV, $m_h = 126.8$~GeV, and
$\Lambda = 1/\sqrt{8\pi G} = 2.435 \times 10^{18}$~GeV.
Also, it is reasonable to put $\tilde{g}(\Lambda)=\lambda_{\chi H}(\Lambda)=0$,
as we mentioned before.
Admittedly, the values are nonperturbative in the UV region, 
as shown in Fig.~\ref{flows}, but
we confirmed that the flows in the IR region are almost unchanged,
even if we vary the compositeness conditions to 
$g^2(\Lambda), y^2(\Lambda), \lambda_\chi (\Lambda) \sim 3\mbox{--}10$.

\begin{figure}[t]
  \begin{center}
  \resizebox{0.35\textheight}{!}
            {\includegraphics{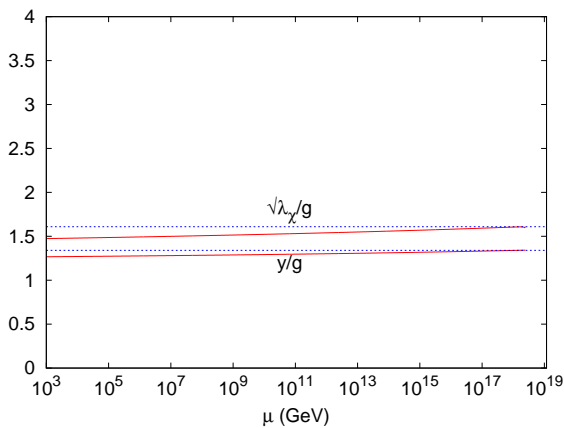}}
  \end{center}
  \caption{The IRFP nature of $\sqrt{\lambda_\chi}/g$ (upper solid line)
  and $y/g$ (lower solid line).
  We took $\Lambda = 1/\sqrt{8\pi G} = 2.435 \times 10^{18}$ GeV.
  The upper and lower dashed lines correspond to the approximate values 
  $\sqrt{2(9+\sqrt{546})}/5 \simeq 1.609$ and $3/\sqrt{5} \simeq 1.342$,
  respectively.
  \label{fig-PRFP}}
\end{figure}

How about the nature of the IRFP?
We show $y/g$ and $\sqrt{\lambda_\chi}/g$ in Fig.~\ref{fig-PRFP}.
These are slightly running, but almost constants.
Numerically, we find the infrared values as
$y/g = 1.27, 1.28, 1.31, 1.32, 1.33$ and 
$\sqrt{\lambda_\chi}/g = 1.47, 1.51, 1.55, 1.57, 1.59$
for $\Lambda=2.435 \times 10^{18}, 10^{15}, 10^{10}, 10^{8}, 10^{6}$~GeV, 
respectively.
Analytical expressions suggest $y/g=3/\sqrt{5}=1.342$
and $\sqrt{\lambda_\chi}/g = \sqrt{2(9+\sqrt{546})}/5=1.609$.

Taking the VEV of $\chi$ as $\VEV{v_\chi} = v_\chi/\sqrt{2}$,
the square of the masses of $\nu_R$, $\chi$ and $Z'$ are
\begin{equation}
 M_{\nu_R}^2 \simeq 2 y^2 v_\chi^2, \quad
 M_\chi^2 \simeq 2\lambda_\chi v_\chi^2, \quad 
 M_{Z'}^2 \simeq 4 g^2v_\chi^2 + g^2 f^2 \, .  
\end{equation}
The IRFP solutions yield the mass relation between $\nu_R$ and $\chi$,
\begin{equation}
  \frac{M_\chi}{M_{\nu_R}} = \frac{\sqrt{\lambda_\chi}}{y} \approx 1.2\,.
\end{equation}
In sharp contrast to the conventional approach for $Z'$,
we have the contribution of the St\"{u}ckelberg mass to $M_{Z'}$,
\begin{equation}
  \Delta \equiv \frac{M_{Z'}^2}{g^2} - 4v_\chi^2 = f^2 > 0 \,.
\end{equation}
If the experiments such as LHC and ILC observe $\Delta > 0$
and confirm $g^2/(4\pi) \gtrsim 0.015$,
it implies the compositeness of $Z'$.

\section{Summary and discussions}

We studied the composite $Z'$ vector boson.
We found that the strong coupling limit of the St\"{u}ckelberg model 
corresponds to the NJL model and that the St\"{u}ckelberg model 
is effectively induced in low energy via the fermion loop from 
the NJL model.
This correspondence can be encoded in terms of the RGE's with
the compositeness conditions.
We showed that the RG flows are determined by the IRFP.
The nature of the IRFP yields the mass ratio,
$M_\chi/M_{\nu_R} = \sqrt{\lambda_\chi}/y \approx 1.2$.
The $Z'$ mass is composed of the VEV of $\chi$ and 
the St\"{u}ckelberg mass.
Owing to this extra mass term, the constraints of the $Z'$ model from 
the precision measurements, the bounds of the direct searches at 
Tevatron, LHC, etc.\cite{Cacciapaglia:2006pk,Erler:2009jh,Salvioni:2009mt}
should be relaxed.  
If $\Delta \equiv M_{Z'}^2/g^2 - 4v_\chi^2 > 0$ is established 
in experiments~\cite{Feldman:2006wb}, 
this might be evidence of the strong dynamics in high energy.

We here considered a scenario that the composite $Z'$ boson 
survives in low energy. From the viewpoint of the naturalness, 
however, the masses of $Z'$, $\chi$ and $\nu_R$ might not be 
so far below the compositeness scale $\Lambda$.
In this case, the see-saw mechanism may work.
We also note that the Higgs potential can be stabilized
by the tree level shift of the Higgs quartic coupling essentially
generated by the $Z'$ loop contribution~\cite{EliasMiro:2012ay}. 
Our approach is unlikely to be applicable to %the composite $W$ and $Z$ in 
the St\"{u}ckelberg extension of the SM~\cite{Kors:2004dx}, 
because the weak and hypercharge gauge couplings are perturbative
up to the Planck scale.
However, the dark matter might be connected with 
the composite $Z'$~\cite{dark-St-Zp}.
We will study several such scenarios elsewhere.


\begin{thebibliography}{99}

\bibitem{Higgs-discover}
%\bibitem{Aad:2012tfa} 
  G.~Aad {\it et al.}  [ATLAS Collaboration],
  %``Observation of a new particle in the search for 
  % the Standard Model Higgs boson with the ATLAS detector at the LHC,''
  Phys. Lett. B {\bf 716}, 1 (2012)
  [arXiv:1207.7214 [hep-ex]];
  %%CITATION = ARXIV:1207.7214;%%
  %1752 citations counted in INSPIRE as of 15 Oct 2013
%\bibitem{Chatrchyan:2012ufa} 
  S.~Chatrchyan {\it et al.}  [CMS Collaboration],
  %``Observation of a new boson at a mass of 125 GeV with 
  % the CMS experiment at the LHC,''
  {\it ibid.} %Phys. Lett. B {\bf 716}, 30 (2012);
  B {\bf 716}, 30 (2012)
  [arXiv:1207.7235 [hep-ex]].
  %%CITATION = ARXIV:1207.7235;%%
  %1727 citations counted in INSPIRE as of 15 Oct 2013    

\bibitem{global-Higgs}
%\bibitem{Ellis:2012rx} 
  J.~Ellis and T.~You,
  %``Global Analysis of Experimental Constraints on a Possible 
  % Higgs-Like Particle with Mass ~ 125 GeV,''
  JHEP {\bf 1206}, 140 (2012)
  [arXiv:1204.0464 [hep-ph]];
  %%CITATION = ARXIV:1204.0464;%%
  %81 citations counted in INSPIRE as of 15 Sep 2014
%\bibitem{Espinosa:2012vu} 
  J.~R.~Espinosa, M.~Muhlleitner, C.~Grojean and M.~Trott,
  %``Probing for Invisible Higgs Decays with Global Fits,''
  JHEP {\bf 1209}, 126 (2012)
  [arXiv:1205.6790 [hep-ph]];
  %%CITATION = ARXIV:1205.6790;%%
  %92 citations counted in INSPIRE as of 15 Sep 2014
%\bibitem{Ellis:2012hz} 
  J.~Ellis and T.~You,
  %``Global Analysis of the Higgs Candidate with Mass ~ 125 GeV,''
  JHEP {\bf 1209}, 123 (2012)
  [arXiv:1207.1693 [hep-ph]];
  %%CITATION = ARXIV:1207.1693;%%
  %117 citations counted in INSPIRE as of 15 Sep 2014
%\bibitem{Cheung:2013kla} 
  K.~Cheung, J.~S.~Lee and P.~Y.~Tseng,
  %``Higgs Precision (Higgcision) Era begins,''
  JHEP {\bf 1305}, 134 (2013)
  [arXiv:1302.3794 [hep-ph]];
  %%CITATION = ARXIV:1302.3794;%%
  %57 citations counted in INSPIRE as of 15 Sep 2014
%\bibitem{Ellis:2013lra} 
  J.~Ellis and T.~You,
  %``Updated Global Analysis of Higgs Couplings,''
  JHEP {\bf 1306}, 103 (2013)
  [arXiv:1303.3879 [hep-ph]];
  %%CITATION = ARXIV:1303.3879;%%
  %121 citations counted in INSPIRE as of 15 Sep 2014
%\bibitem{Belanger:2013xza} 
  G.~Belanger, B.~Dumont, U.~Ellwanger, J.~F.~Gunion and S.~Kraml,
  %``Global fit to Higgs signal strengths and couplings and
  % implications for extended Higgs sectors,''
  Phys. Rev. D {\bf 88}, 075008 (2013)
  [arXiv:1306.2941 [hep-ph]];
  %%CITATION = ARXIV:1306.2941;%%
  %102 citations counted in INSPIRE as of 15 Sep 2014
%\bibitem{Cheung:2014noa} 
  K.~Cheung, J.~S.~Lee and P.~Y.~Tseng,
  %``Higgcision Updates 2014,''
  arXiv:1407.8236 [hep-ph];
  %%CITATION = ARXIV:1407.8236;%%
  %4 citations counted in INSPIRE as of 15 Sep 2014
%\bibitem{Bernon:2014vta} 
  J.~Bernon, B.~Dumont and S.~Kraml,
  %``Status of Higgs couplings after Run-1 of the LHC using Lilith 1.0,''
  arXiv:1409.1588 [hep-ph].
  %%CITATION = ARXIV:1409.1588;%%

\bibitem{stability}
 J.~Elias-Miro, J.~R.~Espinosa, G.~F.~Giudice, G.~Isidori, A.~Riotto and A.~Strumia,
  %``Higgs mass implications on the stability of the electroweak vacuum,''
  Phys. Lett. B {\bf 709}, 222 (2012) 
  [arXiv:1112.3022 [hep-ph]];
  %%CITATION = ARXIV:1112.3022;%%
C.~P.~Burgess, V.~Di Clemente and J.~R.~Espinosa,
  %``Effective operators and vacuum instability as heralds of new physics,''
  JHEP {\bf 0201}, 041 (2002)
  [hep-ph/0201160].
  %%CITATION = HEP-PH/0201160;%%
  D.~Buttazzo, G.~Degrassi, P.~P.~Giardino, G.~F.~Giudice, F.~Sala, A.~Salvio and A.~Strumia,
  %``Investigating the near-criticality of the Higgs boson,''
  JHEP {\bf 1312}, 089 (2013)
  [arXiv:1307.3536].
  %%CITATION = ARXIV:1307.3536;%%
  %54 citations counted in INSPIRE as of 28 Jan 2014

\bibitem{Dimopoulos:1979es} 
  S.~Dimopoulos and L.~Susskind,
  %``Mass Without Scalars,''
  Nucl. Phys. B {\bf 155}, 237 (1979).
  %%CITATION = NUPHA,B155,237;%%
  %1004 citations counted in INSPIRE as of 14 Sep 2014

\bibitem{Eichten:1979ah} 
  E.~Eichten and K.~D.~Lane,
  %``Dynamical Breaking of Weak Interaction Symmetries,''
  Phys. Lett. B {\bf 90}, 125 (1980).
  %%CITATION = PHLTA,B90,125;%%
  %1028 citations counted in INSPIRE as of 14 Sep 2014

\bibitem{Katayama:1962mx} 
  Y.~Katayama, K.~Matumoto, S.~Tanaka and E.~Yamada,
  %``Possible unified models of elementary particles with two neutrinos,''
  Prog. Theor. Phys. {\bf 28}, 675 (1962).
  %%CITATION = PTPKA,28,675;%%
  %146 citations counted in INSPIRE as of 14 Sep 2014

\bibitem{Maki:1962mu} 
  Z.~Maki, M.~Nakagawa and S.~Sakata,
  %``Remarks on the unified model of elementary particles,''
  Prog. Theor. Phys. {\bf 28}, 870 (1962).
  %%CITATION = PTPKA,28,870;%%
  %2450 citations counted in INSPIRE as of 14 Sep 2014

\bibitem{Yasue:1978wg} 
  M.~Yasue,
  %``Quark - Lepton Correspondence and SU(2) X U(1) X U(1) Gauge Model,''
  Prog. Theor. Phys. {\bf 61}, 269 (1979).
  %%CITATION = PTPKA,61,269;%%
  %33 citations counted in INSPIRE as of 14 Sep 2014

\bibitem{Davidson:1978pm} 
  A.~Davidson,
  %``$B^-$l as the Fourth Color, Quark - Lepton Correspondence, 
  % and Natural Masslessness of Neutrinos Within a Generalized Ws Model,''
  Phys. Rev. D {\bf 20}, 776 (1979).
  %%CITATION = PHRVA,D20,776;%%
  %72 citations counted in INSPIRE as of 14 Sep 2014

\bibitem{seesaw}
%\bibitem{Minkowski:1977sc} 
  P.~Minkowski,
  %``mu --> e gamma at a Rate of One Out of 1-Billion Muon Decays?,''
  Phys. Lett. B {\bf 67}, 421 (1977);
  %%CITATION = PHLTA,B67,421;%%
  %1949 citations counted in INSPIRE as of 14 Oct 2014
  T.~Yanagida, in {\it Proceedings of the Workshop on the Unified Theory
  and the Baryon Number in the Universe}, edited by O.Sawada and A. Sugamoto,
  KEK, Tsukuba, Japan, p.95 (1979);
  M.~Gell-Mann, P.~Ramond, and R.~Slansky,
  in {\it Supergravity}, edited by P. Van Nieuwenhuizen and D. Z. Freedman,
  North-Holland, Amsterdam, p.315 (1979).

% \bibitem{seesaw}
% %\bibitem{Minkowski:1977sc} 
%   P.~Minkowski,
%   %``mu --> e gamma at a Rate of One Out of 1-Billion Muon Decays?,''
%   Phys. Lett. B {\bf 67}, 421 (1977):
%   %%CITATION = PHLTA,B67,421;%%
%   %1920 citations counted in INSPIRE as of 14 Sep 2014
%   T. Yanagida, Proceedings of the Workshop on Unified
%   Theories and Baryon Number in the Universe, Tsukuba, Japan 1979,
%   eds. A. Sawada and A. Sugamoto;
%   M. Gell-Mann, P. Ramond and R. Slansky, 
%   Proceedings of the Supergravity Stony Brook Workshop, New York, 
%   eds. P. Van Nieuwenhuizen and D. Freedman.

\bibitem{Langacker:2008yv} 
   P.~Langacker,
   %``The Physics of Heavy $Z^\prime$ Gauge Bosons,''
   Rev. Mod. Phys. {\bf 81}, 1199 (2009)
   [arXiv:0801.1345 [hep-ph]].
   %%CITATION = ARXIV:0801.1345;%%
   %466 citations counted in INSPIRE as of 06 Jan 2014

\bibitem{Leike:1998wr} 
  A.~Leike,
  %``The Phenomenology of extra neutral gauge bosons,''
  Phys. Rept. {\bf 317}, 143 (1999)
  [hep-ph/9805494].
  %%CITATION = HEP-PH/9805494;%%

\bibitem{Basso:2010jm} 
  L.~Basso, S.~Moretti and G.~M.~Pruna,
  %``A Renormalisation Group Equation Study of the Scalar Sector of
  % the Minimal B-L Extension of the Standard Model,''
  Phys. Rev. D {\bf 82}, 055018 (2010)
  [arXiv:1004.3039 [hep-ph]].
  %%CITATION = ARXIV:1004.3039;%%
  %19 citations counted in INSPIRE as of 26 Feb 2013

\bibitem{Terazawa:1976xx} 
  H.~Terazawa, K.~Akama and Y.~Chikashige,
  %``Unified Model of the Nambu-Jona-Lasinio Type for 
  % All Elementary Particle Forces,''
  Phys. Rev.D {\bf 15}, 480 (1977);
  %%CITATION = PHRVA,D15,480;%%
  %409 citations counted in INSPIRE as of 19 Sep 2014
%\bibitem{Terazawa:1979pj} 
  H.~Terazawa,
  %``Subquark Model of Leptons and Quarks,''
  Phys. Rev. D {\bf 22}, 184 (1980).
  %%CITATION = PHRVA,D22,184;%%
  %186 citations counted in INSPIRE as of 19 Sep 2014

\bibitem{Stueckelberg:1900zz} 
  E.~C.~G.~Stueckelberg,
  %``Interaction energy in electrodynamics and in the field theory of 
  % nuclear forces,''
  Helv. Phys. Acta {\bf 11}, 225 (1938).
  %%CITATION = HPACA,11,225;%%
  %206 citations counted in INSPIRE as of 14 Sep 2014

\bibitem{Ruegg:2003ps} 
  See for review, H.~Ruegg and M.~Ruiz-Altaba,
  %``The Stueckelberg field,''
  Int. J. Mod. Phys. A {\bf 19}, 3265 (2004)
  [hep-th/0304245].
  %%CITATION = HEP-TH/0304245;%%
  %130 citations counted in INSPIRE as of 14 Sep 2014

\bibitem{Bardeen:1989ds} 
  W.~A.~Bardeen, C.~T.~Hill and M.~Lindner,
  %``Minimal Dynamical Symmetry Breaking of the Standard Model,''
  Phys. Rev. D {\bf 41}, 1647 (1990).
  %%CITATION = PHRVA,D41,1647;%%
  %1012 citations counted in INSPIRE as of 13 Sep 2014

\bibitem{Miransky:1988xi}
  V.~A.~Miransky, M.~Tanabashi and K.~Yamawaki,
  %``Dynamical Electroweak Symmetry Breaking with Large Anomalous Dimension and
  %t Quark Condensate,''
  Phys. Lett. B {\bf 221}, 177 (1989);
  %%CITATION = PHLTA,B221,177;%%
  %``Is the t Quark Responsible for the Mass of W and Z Bosons?,''
  Mod. Phys. Lett. A {\bf 4}, 1043 (1989).
  %%CITATION = MPLAE,A4,1043;%%

\bibitem{Nambu} 
  Y. Nambu, Enrico Fermi Institute Report No. 89-08, 1989;
  in {\it Proceedings of the 1988 Kazimierz Workshop}, eds. Z. Ajduk et al. 
  (World Scientific Publishing Co., Singapore, 1989).

\bibitem{Hill:2002ap}
  For comprehensive reviews, see, e.g.,
  C.~T.~Hill and E.~H.~Simmons,
  %``Strong dynamics and electroweak symmetry breaking,''
  Phys. Rept. {\bf 381}, 235 (2003)
  [Erratum-{\it ibid.} {\bf 390}, 553 (2004)];
  %%CITATION = PRPLC,381,235;%%
%\bibitem{Cvetic:1997eb} 
  G.~Cvetic,
  %``Top quark condensation,''
  Rev. Mod. Phys. {\bf 71}, 513 (1999)
  [hep-ph/9702381].
  %%CITATION = HEP-PH/9702381;%%
  %170 citations counted in INSPIRE as of 17 Sep 2014

\bibitem{Terazawa:1990mz} 
  H.~Terazawa,
  %``t quark mass predicted from a sum rule for lepton and quark masses,''
  Phys. Rev. D {\bf 22}, 2921 (1980)
  [Erratum-ibid. D {\bf 41}, 3541 (1990)].
  %%CITATION = PHRVA,D22,2921;%%
  %39 citations counted in INSPIRE as of 19 Sep 2014

\bibitem{Pendleton:1980as} 
  B.~Pendleton and G.~G.~Ross,
  %``Mass and Mixing Angle Predictions from Infrared Fixed Points,''
  Phys. Lett. B {\bf 98}, 291 (1981).
  %%CITATION = PHLTA,B98,291;%%
  %411 citations counted in INSPIRE as of 21 Sep 2013

\bibitem{EliasMiro:2012ay} 
  J.~Elias-Miro, J.~R.~Espinosa, G.~F.~Giudice, H.~M.~Lee and A.~Strumia,
  %``Stabilization of the Electroweak Vacuum by a Scalar Threshold Effect,''
  JHEP {\bf 1206}, 031 (2012)
  [arXiv:1203.0237 [hep-ph]].
  %%CITATION = ARXIV:1203.0237;%%
  %95 citations counted in INSPIRE as of 14 Oct 2014

\bibitem{Bando:1987br} 
  M.~Bando, T.~Kugo and K.~Yamawaki,
  %``Nonlinear Realization and Hidden Local Symmetries,''
  Phys. Rept. {\bf 164}, 217 (1988).
  %%CITATION = PRPLC,164,217;%%
  %892 citations counted in INSPIRE as of 13 Sep 2014

\bibitem{Harada:2003jx} 
  M.~Harada and K.~Yamawaki,
  %``Hidden local symmetry at loop: A New perspective of %
  % composite gauge boson and chiral phase transition,''
  Phys. Rept. {\bf 381}, 1 (2003)
  [hep-ph/0302103].
  %%CITATION = HEP-PH/0302103;%%
  %360 citations counted in INSPIRE as of 13 Sep 2014

\bibitem{Iso:2009ss} 
  S.~Iso, N.~Okada and Y.~Orikasa,
  %``Classically conformal $B^-$ L extended Standard Model,''
  Phys. Lett. B {\bf 676}, 81 (2009)
  [arXiv:0902.4050 [hep-ph]];
  %%CITATION = ARXIV:0902.4050;%%
  %34 citations counted in INSPIRE as of 05 Apr 2013
%\bibitem{Iso:2009nw} 
  %S.~Iso, N.~Okada and Y.~Orikasa,
  %``The minimal B-L model naturally realized at TeV scale,''
  Phys. Rev. D {\bf 80}, 115007 (2009)
  [arXiv:0909.0128 [hep-ph]];
  %%CITATION = ARXIV:0909.0128;%%
  %20 citations counted in INSPIRE as of 05 Apr 2013
%\bibitem{Iso:2012jn} 
  S.~Iso and Y.~Orikasa,
  %``TeV Scale B-L model with a flat Higgs potential at the Planck
  % scale - in view of the hierarchy problem -,''
  PTEP {\bf 2013}, 023B08 (2013)
  [arXiv:1210.2848 [hep-ph]].
  %%CITATION = ARXIV:1210.2848;%%
  %7 citations counted in INSPIRE as of 05 Apr 2013

\bibitem{Hashimoto:2014ela} 
  M.~Hashimoto, S.~Iso and Y.~Orikasa,
  %``Radiative Symmetry Breaking from Flat Potential in various U(1)' models,''
  Phys. Rev. D {\bf 89}, 056010 (2014)
  [arXiv:1401.5944 [hep-ph]].
  %%CITATION = ARXIV:1401.5944;%%
  %10 citations counted in INSPIRE as of 10 Sep 2014

\bibitem{Hashimoto:2013hta} 
  M.~Hashimoto, S.~Iso and Y.~Orikasa,
  %``Radiative symmetry breaking at the Fermi scale and flat potential 
  % at the Planck scale,''
  Phys. Rev. D {\bf 89}, 016019 (2014)
  [arXiv:1310.4304 [hep-ph]].
  %%CITATION = ARXIV:1310.4304;%%
  %13 citations counted in INSPIRE as of 10 Sep 2014

\bibitem{Cacciapaglia:2006pk} 
  G.~Cacciapaglia, C.~Csaki, G.~Marandella and A.~Strumia,
  %``The Minimal Set of Electroweak Precision Parameters,''
  Phys. Rev. D {\bf 74}, 033011 (2006)
  [hep-ph/0604111].
  %%CITATION = HEP-PH/0604111;%%
  %72 citations counted in INSPIRE as of 20 Mar 2013

\bibitem{Erler:2009jh} 
  J.~Erler, P.~Langacker, S.~Munir and E.~Rojas,
  %``Improved Constraints on Z-prime Bosons from Electroweak Precision Data,''
  JHEP {\bf 0908}, 017 (2009)
  [arXiv:0906.2435 [hep-ph]].
  %%CITATION = ARXIV:0906.2435;%%
  %94 citations counted in INSPIRE as of 27 Feb 2013

\bibitem{Salvioni:2009mt} 
  E.~Salvioni, G.~Villadoro and F.~Zwirner,
  %``Minimal Z-prime models: Present bounds and early LHC reach,''
  JHEP {\bf 0911}, 068 (2009)
  [arXiv:0909.1320 [hep-ph]].
  %%CITATION = ARXIV:0909.1320;%%
  %53 citations counted in INSPIRE as of 08 Mar 2013

\bibitem{Feldman:2006wb} 
  D.~Feldman, Z.~Liu and P.~Nath,
  %``The Stueckelberg $Z$ Prime at the LHC: Discovery Potential, 
  % Signature Spaces and Model Discrimination,''
  JHEP {\bf 0611}, 007 (2006)
  [hep-ph/0606294].
  %%CITATION = HEP-PH/0606294;%%
  %73 citations counted in INSPIRE as of 17 Sep 2014

\bibitem{Kors:2004dx} 
  B.~Kors and P.~Nath,
  %``A Stueckelberg extension of the standard model,''
  Phys. Lett. B {\bf 586}, 366 (2004)
  [hep-ph/0402047];
  %%CITATION = HEP-PH/0402047;%%
  %125 citations counted in INSPIRE as of 17 Sep 2014
%\bibitem{Kors:2005uz} 
%  B.~Kors and P.~Nath,
  %``Aspects of the Stueckelberg extension,''
  JHEP {\bf 0507}, 069 (2005)
  [hep-ph/0503208].
  %%CITATION = HEP-PH/0503208;%%
  %99 citations counted in INSPIRE as of 17 Sep 2014

\bibitem{dark-St-Zp}
%\bibitem{Cheung:2007ut} 
  K.~Cheung and T.~C.~Yuan,
  %``Hidden fermion as milli-charged dark matter in Stueckelberg 
  % Z- prime model,''
  JHEP {\bf 0703}, 120 (2007)
  [hep-ph/0701107];
  %%CITATION = HEP-PH/0701107;%%
  %59 citations counted in INSPIRE as of 17 Sep 2014
%\bibitem{Feldman:2007wj} 
  D.~Feldman, Z.~Liu and P.~Nath,
  %``The Stueckelberg Z-prime Extension with Kinetic Mixing and 
  % Milli-Charged Dark Matter From the Hidden Sector,''
  Phys. Rev. D {\bf 75}, 115001 (2007)
  [hep-ph/0702123 [HEP-PH]];
  %%CITATION = HEP-PH/0702123;%%
  %157 citations counted in INSPIRE as of 17 Sep 2014
%\bibitem{Cheung:2010az} 
  K.~Cheung, K.~H.~Tsao and T.~C.~Yuan,
  %``Hidden Sector Dirac Dark Matter, Stueckelberg Z' Model and 
  % the CDMS and XENON Experiments,''
  arXiv:1003.4611 [hep-ph];
  %%CITATION = ARXIV:1003.4611;%%
  %8 citations counted in INSPIRE as of 17 Sep 2014
%\bibitem{Feng:2014cla} 
  W.~Z.~Feng, G.~Shiu, P.~Soler and F.~Ye,
  %``Building a Stueckelberg portal,''
  JHEP {\bf 1405}, 065 (2014)
  [arXiv:1401.5890 [hep-ph]];
  %%CITATION = ARXIV:1401.5890;%%
  %4 citations counted in INSPIRE as of 17 Sep 2014
%\bibitem{Santos:2014xka} 
  A.~L.~Dos Santos and D.~Hadjimichef,
  %``Astrophysical aspects of milli-charged dark matter in 
  % a Higgs-Stueckelberg model,''
  arXiv:1405.4282 [hep-ph].
  %%CITATION = ARXIV:1405.4282;%%
  %1 citations counted in INSPIRE as of 17 Sep 2014

\end{thebibliography}
\end{document}